\documentclass[12pt]{article}
\usepackage{graphicx,verbatim,array,multicol,palatino,amsfonts,amsmath,subfigure, psfrag}
\usepackage{chicago}
\usepackage{color}

\usepackage{a4,epsf}
\usepackage{latexsym,amsfonts,amsmath,amssymb,epsfig,graphics}

\usepackage{mathtools}
\usepackage{amsmath}

\usepackage{setspace}

\def\bfepsilon{{\bf \epsilon}}

\DeclareMathOperator*{\argmin}{arg\,min}

\title{On Bayesian quantile regression curve fitting via auxiliary variables}
\author{
J.-L. Dortet-Bernadet \\
{\small Institut de Recherche Math\'ematique Avanc\'ee, UMR 7501 CNRS}\\
{\small Universit\'e Louis Pasteur, Strasbourg, France} \\
\and
Y. Fan\\
{\small School of Mathematics and Statistics}\\
{\small University of New South Wales, Sydney 2052, Australia}\\
}

\begin{document}

\maketitle

\abstract

Quantile regression has received increased attention in the statistics community in recent years.  
This article adapts an auxiliary variable method, commonly used in Bayesian variable selection for mean regression models, to the fitting of quantile regression curves. We focus on the fitting of regression splines, with unknown number and location of knots. We  provide an efficient algorithm with Metropolis-Hastings updates whose 
tuning is fully automated. 
The method is tested on simulated and real examples and its extension to  additive  models is described.
Finally we propose a simple postprocessing procedure to deal with the problem of the crossing of multiple separately estimated quantile curves. 


\section{Introduction}
Quantile regression has been recognized in recent years as a robust statistical procedure that offers a powerful  alternative to the ordinary mean regression, 
especially when the data contains large outliers or when the response variable has  a skewed or  multimodal conditional distribution.
Given a fixed probability $p$, $0<p<1$, let the model corresponding to the $p$-th quantile regression curve be given by
\begin{eqnarray*}
Y_i|x_1,...,x_n & \sim & f_p(x_i)+\epsilon_i, \quad i=1,\ldots,n
\end{eqnarray*}
where $\epsilon_1,...,\epsilon_n$ are independent draws from a noise distribution whose $p$-th quantile is $0$, {\it i.e.} 
$\mathbb{P}(\epsilon\leq 0)=p$. 
Under this model the $p$-th quantile of the conditional distribution of $Y$ given $\{X=x\}$ is  given by some smooth function $f_p(x)$. If the distribution of 
the noise is left unspecified then the estimation of $f_p$ is typically carried out by solving the minimization problem, for a given class ${\cal F}$ of curves,
\begin{eqnarray}
\argmin_{f_p  \in {\cal F}} \sum_{i=1}^n \rho_p(y_i-f(x_i))
\label{checkfunction}
\end{eqnarray}
where the so-called  "check function'' $\rho_p(.)$ is given by $\rho_p(\epsilon)=p \epsilon $ if $\epsilon\geq 0$ and $\rho_p(\epsilon)=(p-1) \epsilon $ otherwise (see  \citeNP{KoenkerBasset1978}). 
To define a likelihood function, one usually assumes that the noise distribution is an asymetric Laplace distribution so that 
the maximum likelihood estimate corresponds to the solution of the minimization problem  (see \citeNP{Koenker1999}). 
See  {\it e.g.} \shortciteN{Yu2003} or  \shortciteN{Koenker2005}  for a review on quantile regression  
and  \shortciteN{Geraci2007} for quantile regression with longitudinal data. 
References on Bayesian treatments of the subject include \shortciteN{Tsionas2003} for inference on a  single quantile, 
\shortciteN{YuMoyeed2001} for quantile regression with a  random walk Metropolis-Hastings
algorithm and  \shortciteN{Yu2002} for 
quantile regression with a reversible jump MCMC sampler (RJMCMC, \shortciteNP{green95}). More recently \shortciteN{YueRue2011}  considers additive mixed regression models and inference with either 
MCMC sampling or the integrated nested Laplace approximation (INLA, \shortciteNP{Rue2009}) and \shortciteN{KozumiKobayashi2011} proposed quantile regression with a Gibbs sampler.

In this article, we are interested in the case where the curve $f_p$ is 
modeled with spline functions of a given degree, $P\geq 1$, so that,
\begin{equation}\label{eqn:spline}
f_p(x) =  \alpha_0 + \sum_{j=1}^P \alpha_jx^j + \sum_{k=1}^K \eta_k(x-\gamma_k)_+^P
\end{equation}
where $z_+=\max(0,z)$ and where $\gamma_k, k=1,\ldots,K $ represent the locations of  $K$ knot points (see \citeNP{hastie+t90}). 
Typically,  the degree $P$ is set to equal 3, since  cubic splines are known to approximate locally smooth functions arbitrarily 
well.
\shortciteN{ChenYu2009} provides a Bayesian inference on this model, 
where the number of knots and their location are automatically selected. Their method relies on a  RJMCMC algorithm which, under the prior specifications they use, needs to compute  an approximation of the ratio of  marginal likelihoods. 
For fitting of quantile smoothing splines  see  \shortciteN{Koenker1994} and \shortciteN{He1999} and for a Bayesian inference with natural cubic splines see 
 \shortciteN{Thompson2010}.

We propose  here an alternative strategy that avoids the use of the  RJMCMC sampler which can often be difficult to tune (see \shortciteNP{fans10}  for a review) 
and that does not rely on approximations to simplify computations. 
Recognising that a Bayesian variable selection technique  (e.g. \shortciteNP{george+M93}) can be used  for inference on a curve  
(e.g. \shortciteNP{smith+kohn96},   \shortciteNP{FanSplines10})
we use an auxiliary variable approach which makes possible, under appropriate prior specifications, a Metropolis-Hastings within Gibbs sampler. 
The proposed MCMC sampler  is  easy to implement and fully automated. In particular it incorporates an algorithm which  automatically tunes the scaling parameters used in our Random-walk Metropolis-Hastings algorithm. 

 In Section \ref{modelCode} we present the model and  the prior specifications, then we describe how inference is carried out with a MCMC sampler.  
 We  apply the method on several datasets in Section \ref{examples}.
 In Section \ref{addMod}  we consider  quantile curve regression for additive models. 
 Finally,  in  Section \ref{crossing},   we discuss  the problem of crossing quantile curves and propose a simple postprocessing procedure to reweight the MCMC samples from separately estimated quantile curves.


\section{\label{modelCode}Quantile regression with splines}
For some $0<p<1$, and given paired observations   $(x_1, y_1), \ldots, (x_n, y_n)$,  we are interested in fitting the $p$-th quantile regression model
\begin{eqnarray}
Y_i|x_1,...,x_n & \sim & f_p(x_i)+\epsilon_i, \quad i=1,\ldots,n
\label{model2}
\end{eqnarray}
where $\epsilon_1,...,\epsilon_n$ are independent draws from the asymetric Laplace distribution
\begin{eqnarray}
d_{ ALp(0,\sigma)}(\epsilon) & = &  \frac{p(1-p)}{\sigma} \exp \left[ -\frac{1}{\sigma} \rho_p(\epsilon)\} \right]
\label{LaplaceAsym}
\end{eqnarray}
 for an unknown scale parameter $\sigma > 0$.
Under this model the $p$-th quantile of the conditional distribution of $Y$ given $\{X=x\}$ is  $f_p(x)$. The asymetric Laplace distribution has been adopted in many papers, see for example \citeN{Koenker1999},  \shortciteN{YuMoyeed2001},   \shortciteN{Tsionas2003}, \shortciteN{ChenYu2009}, \shortciteN{YueRue2011} or 
\shortciteN{KozumiKobayashi2011} .  
Under the asymmetric Laplace distribution, given $\sigma$, the function $f_p$  maximizing the likelihood corresponding to model (\ref{model2}) is also the solution of the minimization problem in Equation (\ref{checkfunction}). The scale parameter 
 $\sigma$ that takes into account the variability of the observations is considered as a nuisance parameter.

We consider hereafter that the curve $f_p$ is modeled with spline functions of a given degree $P > 0$, in the form of Equation (\ref{eqn:spline}).
Under this representation, fitting 
the curve consists of estimating the number of knots $K$, the knot locations 
$\gamma_k, k=1,\ldots,K$, and the corresponding regression coefficients $\alpha_j$, $j=0,\ldots ,P$ and $\eta_k$, $k=1,\ldots,K$.
If $\gamma_k$, $k=1,...,K_{max}$, where $K_{max}$ represents the (known) maximum number of potential knots,  model (\ref{model2}) can be written as the linear model
\begin{equation}\label{eqn:varsele}
Y=X_{\gamma}\beta + \bfepsilon
\end{equation}
where $Y=(y_1,\ldots,y_n)'$, $\beta=(\alpha_0,\alpha_1,\ldots,\alpha_P,\eta_1,\ldots, \eta_{K_{max}})'$, $\bfepsilon=(\epsilon_1, \ldots, \epsilon_n)'$, 
with design matrix 
\begin{eqnarray}
X_{\gamma} & = & ({\bf 1}_n,{\bf x}, \ldots ,{\bf x}^P, ({\bf x}-{\bf 1}_n\gamma_1)^P_+, \ldots , ({\bf x}-{\bf 1}_n \gamma_{K_{max}})^P_+)
\label{designMatrix}
\end{eqnarray}
where ${\bf x}=(x_1,\ldots ,x_n)'$ and where ${\bf 1}_n=(1,\ldots,1)'$ denotes the unit vector of size $n$.

\subsection{The model and prior assumptions}
We adopt an auxiliary variable approach for the spline regression model by introducing a vector of binary
indicator variables $z_k, k=1,\ldots, K_{max}$, 
$$z_k =\left\{\begin{array}{ll}
1 &\quad \mbox{if there is a knot point } \gamma_k \mbox{ in the interval } 
I_k \mbox{ and } \eta_k\neq 0\\
0 &\quad \mbox{if there is no knot point in the interval }I_k \mbox{ and } \eta_k = 0
\end{array}\right.$$
where $\eta_k$ denotes the spline coefficients in model (\ref{eqn:varsele}), and the intervals $I_k$ are defined on the range of the $x_i$'s.  Each interval $I_k$  contains at most one knot with unknown location $\gamma_k$. 
In practice, such intervals can be defined by either using prior information on regions where a knot is suspected or, in the absence of such prior information, an equal partition of the range
may be adopted. We denote the vector $(\gamma_1, \ldots , \gamma_{K_{max}})'$  
by $\gamma$  and  consider the Uniform distributions on the interval as the prior distribution on $\gamma$. Each possible value for  $\gamma$ gives a model of the form (\ref{eqn:varsele}). Let $X_{z, \gamma}$ denotes the matrix constructed with the columns of $X_{\gamma}$ corresponding to non-zero entries in $z$, and let $\beta_{z,\gamma}$ denotes the vector of corresponding regression coefficients.

A desirable feature of the asymetric Laplace distribution  is that it can be decomposed as a scale mixture of normals (see {\it e.g.} \citeNP{Tsionas2003}, \citeNP{YueRue2011} or \shortciteNP{KozumiKobayashi2011})
\begin{eqnarray*} 
\epsilon|w & \sim & {\cal N}\left( \frac{(1-2p)w}{p(1-p)},\frac{2\sigma w}{p(1-p)} \right) ,\\
w & \sim & {\cal E}xp(1/\sigma), \nonumber
\end{eqnarray*}
where ${\cal E}xp(1/\sigma)$ denotes the exponential distribution with mean $\sigma$. If $w_i$, $i=1,\ldots,n$ denote  the variable $w$ 
associated with  each $\epsilon_i$, the conditional distribution of $Y$ given  $W$,  the diagonal matrix with entries $w_i$, $i=1,\ldots,n$, is 
\begin{eqnarray}
f(Y| X_{z, \gamma},\beta_{z,\gamma},z, \sigma, \gamma ,W)  & = & {\cal N} \left( X_{z,\gamma} \beta_{z,\gamma}+\frac{(1-2p)}{p(1-p)}W {\bf 1}_n,\frac{2 \sigma}{p(1-p)} W \right).
\label{condYquantilebis}
\end{eqnarray}
Conditional on $W$ we use the following decomposition of the joint prior distribution of the unknown parameters
$$
\pi( \beta_{z,\gamma},z, \sigma, \gamma|W )=\pi_{\beta_{z,\gamma}}(\beta_{z,\gamma}|z, \sigma, \gamma,W)\pi_{\sigma}(\sigma)\pi_z(z)\pi_{\gamma}(\gamma),
$$
where we set
\begin{eqnarray}
\pi_{\beta_{z,\gamma}}(\beta_{z,\gamma}|z, \sigma, \gamma,W) & = &  {\cal N} \left( 0, \frac{2\sigma}{p(1-p)} c(X'_{z,\gamma}W^{-1}X_{z,\gamma} )^{-1} \right).
\label{unitinfoQuant}
\end{eqnarray}
This conditional prior for $\beta_{z,\gamma}$, related to $g$-priors (\shortciteNP{zellner86}), has the advantage of conjugacy in the case of 
normal errors, in which case the regression and variance parameters can be analytically integrated out. 

Different choices for the parameter $c$ have been proposed in the literature for mean regression problems. The case $c=n$, where $n$ is the sample size, corresponds to the unit information prior which was used by \shortciteN{dimatteo+gk01}, a default choice that works well in 
practice in Bayesian variable selection problems with large sample sizes. \shortciteN{smith+kohn96} recommend values of $c$ in the range $10 \leq c \leq 1000$ for the problems they considered. 
Here including an adaptive scale parameter $c$, and treating it as another parameter  was more satisfactory than using a fixed one. 
Thus we include a hyper-prior for $c$, following  {\it e.g.}   \shortciteN{Leslie2007}, we use a diffuse prior ${\cal IG}(1,2n)$ with a mode at $n$
\begin{eqnarray*}
\pi(c) & \propto & c^{-2} \exp\{-2n/c \}.
\end{eqnarray*}
See  \shortciteN{Liang2008}  for more discussion about the choice of a prior  distribution on  the parameter $c$.

For the variance parameter, we use the standard uninformative prior $\pi_{\sigma}(\sigma) \propto 1/\sigma$. 
Finally, we need to define the prior distribution for $z$, we consider  the decomposition of this prior given by
$$
\pi_z(z)=\pi(z\mid |z|)\pi(|z|) 
$$
where $|z|=\sum_{k=1}^{K_{max}}z_k$ is the number of non-zero entries in $z$, {\it i.e.} the number of knots that are used in the corresponding model. 
We use for this term a Poisson  distribution with mean $\lambda$  that is right-truncated at a specified maximum value, $L$. We assume also that, given 
this quantity, all possible configurations for $z$ have equal probabilities, so that
\begin{eqnarray*}
\pi_z(z) & \propto & \frac{\lambda^{|z|}}{|z|!} I_{\{|z|\leq L\}}.
\end{eqnarray*}
The parameters  $\beta_{z,\gamma}$ and $\sigma$ can be integrated out of the  
full joint posterior distribution $\pi( \beta_{z,\gamma},z, \sigma, \gamma,W,c|Y)$  
and we get 
\begin{eqnarray}
\pi(z, \gamma,W,c |Y) & \propto & \frac{\pi(c)  \pi_z(z)\pi_{\gamma}(\gamma)   }{  {  \sqrt{ \prod_{i=1}^n w_i }  }     (c+1)^{(|z|+P+1)/2}}  \left\{ \frac{p(1-p)}{4}S_{z,\gamma,W,c}(Y)  +\sum_{i=1}^n w_i \right\}^{-3n/2} 
\label{posteriorintegQuant}
\end{eqnarray}
where 
\begin{eqnarray*}
S_{z,\gamma,W,c}(Y) & = & Y_{(W)}'W^{-1}Y_{(W)}-\frac{c}{c+1}Y_{(W)}'W^{-1}X_{z,\gamma}(X'_{z,\gamma}W^{-1} X_{z,\gamma})^{-1}X'_{z,\gamma}W^{-1}Y_{(W)}\nonumber \\
\end{eqnarray*}
and where
\begin{eqnarray*}
Y_{(W)} & = & Y-\frac{(1-2p)}{p(1-p)}W {\bf 1}_n.
\end{eqnarray*}
Details of the  marginal posterior are  given in Appendix A.

\subsection{Inference on the posterior distribution}\label{sec:alg}
An MCMC sampler is used for the inference on the model. Based on the posterior distribution (\ref{posteriorintegQuant}), for each $t^{th}$ iteration of the MCMC update, $t=1,\ldots, T$, perform the following successive updates for $z$, $\gamma$, $W$ and $c$:
\begin{itemize}
\item {\bf Update $z$}. This update involves two types of moves; with probability 0.5 we propose
an add/delete step, otherwise a swap step is proposed. 
Specifically, the two move steps involve
\begin{itemize}
\item add/delete: randomly select a $z_k$ and propose to change its value;
\item swap: randomly select two values $z_i$ and $z_j$, 
and propose to exchange their values. 
\end{itemize}
In both cases, proposed moves from current value $z$ to proposed value 
$z'$ are accepted with the usual Metropolis-Hastings acceptance probability
$$
\alpha(z, z') = \mbox{min} \left\{1, \frac{\pi(z', \gamma,W,c|Y) q(z', z)}{\pi(z, \gamma,W,c|Y) q(z, z')} \right\} 
$$
where $q(z, z')$ is the probability of proposing the new value $z'$ given the current value $z$. 

\item {\bf Update $\gamma$}. For each $k=1,\ldots, K_{max}$, we differentiate the cases when $z_k=0$ and when $z_k=1$:  
\begin{itemize}
\item if $z_k=0$ then $\gamma_k$ is updated according to its prior distribution, {\it i.e.} a  Uniform distribution on $I_k$; 
\item  
if $z_k=1$, $\gamma_k$ is updated to a new value $\gamma^{'}_k$, according to the posterior distribution
$$
\pi(\gamma_k|\gamma_{j\neq k},z,Y,W,c)\propto   \left\{ \frac{p(1-p)}{4}S_{z,\gamma,W,c}(Y)  +\sum_{i=1}^n w_i \right\}^{-3n/2}        \pi_{\gamma}(\gamma).
$$
\end{itemize}
An  independence Metropolis-Hastings step can be used for this last type of updating, using the prior on $\gamma_k$ as a proposal, with the 
corresponding acceptance probability given by
$$
\alpha(\gamma_k, \gamma^{'}_k) = \mbox{min} \left\{1, \frac{\pi(\gamma^{'}_k|\gamma_{j\neq k},z,Y,W,c)}{\pi(\gamma_k|\gamma_{j\neq k},z,Y,W,c)}
\right\} .$$

\item {\bf Update $W$}. Each $w_i, i=1,\ldots,n$ has conditional posterior distribution 
\begin{eqnarray*}
\pi(w_i| w_{i\neq j}\gamma,z,c,Y) & \propto & \frac{1}{\prod_{i=1}^n \sqrt{w_i}} \left\{ \frac{p(1-p)}{4}S_{z,\gamma,W,c}(Y)  +\sum_{i=1}^n w_i \right\}^{-3n/2}. 
\end{eqnarray*}
We use a Random-walk Metropolis-Hastings proposal to update each $w_i$.  We consider as proposal distribution a normal distribution 
$q(w_i, .)=N(w_i, \sigma^2_{i})$  with mean $w_i$ and variance $\sigma^2_{i}$.   
We sample $w'_i \sim q(w_i, .)$, then the proposed value $w'_i$ is accepted with probability
$$
\alpha(w_i,w'_i)=\min\left\{1, \frac{\pi(w'_i| w_{i\neq j}\gamma,z,c,Y) }{\pi(w_i| w_{i\neq j}\gamma,z,c,Y) }\right\}.
$$
The tuning parameters $\sigma^2_{i},  i=1,\ldots, n$ are optimally obtained automatically, prior to starting the main part of MCMC, see Appendix B.

\item {\bf Update $c$}.  The parameter $c$ has conditional distribution
\begin{eqnarray*}
\pi(c|z,\gamma,W,Y) & \propto &  \frac{\pi(c)    }{    (c+1)^{(|z|+P+1)/2}}  \left\{ \frac{p(1-p)}{4}S_{z,\gamma,W,c}(Y)  +\sum_{i=1}^n w_i \right\}^{-3n/2}  .
\end{eqnarray*}
We use a Random-walk Metropolis-Hastings proposal to update $c$. We sample $c'\sim q(c, .)=N(c, \sigma_*^2)$ then accept the proposed value with acceptance probability
$$
\alpha(c, c')=\min\left\{1, \frac{\pi(c'| z,\gamma,W,Y) }{\pi(c| z,\gamma,W,Y) }\right\}.
$$
The tuning parameter $\sigma^2_*$ is also obtained via the algorithm in Appendix B.
 
\end{itemize}

\noindent Note that when the sample size $n$ is large, the number of parameters in the Update $W$ step becomes large  and correspondingly manual tuning of the scale parameters $\sigma^2_{i}$ in the Gaussian Random-Walk Metropolis-Hastings sampler becomes infeasible. One strategy to automate the sampler is to use a slice sampler (see \shortciteNP{neal03}). But the additional evaluations of the posterior function makes this algorithm much more computationally intensive.
In this article we use the algorithm of \shortciteN{garthwaitefs11} that  automatically tunes the scaling parameters $\sigma_i^2$  and obtains an optimal over all acceptance rate of  $p^*=0.44$ (\shortciteNP{roberts+r01}) for these univariate updates.  See Appendix B for a description of the algorithm used for tuning.

Once a converged MCMC sample $\{(z^{(t)},\gamma^{(t)},W^{(t)},c^{(t)} )\}_{t=1,...,T}$ is obtained it is possible to estimate the curve $f_p(x)$ by a Bayesian model averaging approach (BMA). 
The posterior expectation for $\beta$ given $z$, $\gamma$, $W$ and $c$ is 
\begin{eqnarray}
\label{postBetaQuant}
\mathbb{E}(\beta_{z,\gamma}|z,\gamma,W,Y,c) & = & \frac{c}{c+1}(X'_{z,\gamma}W^{-1} X_{z,\gamma})^{-1}X'_{z,\gamma}W^{-1} Y_{(W)}.
\end{eqnarray}
Thus  an estimate for  $f_p(x)$ can be obtained by
\begin{eqnarray*}
\hat{f}_p^{BMA}(x) & = & \frac{1}{T} \sum_{t=1}^T    X_{z^t,\gamma^t}    \frac{c^{(t)}}{c^{(t)}+1}  (X'_{z^t,\gamma^t}(W^t)^{-1} X_{z^t,\gamma^t})^{-1}X'_{z^t,\gamma^t}(W^t)^{-1}Y_{(W^t)}.
\end{eqnarray*}
Another possibility to estimate the curve $f_p(x)$ is to use the maximum a posteriori (MAP) estimate for $( z, \gamma ,W ,c  )$
$$ 
(\hat{z}, \hat{\gamma}, \hat{W},\hat{c}) = \underset{1\leq t \leq T}{\mbox{argmax }} \pi(z^{(t)}, \gamma^{(t)}, W^{(t)},c^{(t)}|Y),
$$
then calculate the corresponding  curve estimate via
\begin{equation*}
\hat{f}_p^{MAP} (x) =  \frac{\hat{c}}{\hat{c}+1} X_{\hat{z}, \hat{\gamma} }  (X'_{\hat{z},\hat{\gamma}} \hat{W}^{-1} X_{\hat{z},\hat{\gamma}})^{-1}X'_{\hat{z},\hat{\gamma}}\hat{W}^{-1} Y_{(\hat{W})}  .
\end{equation*}

\section{\label{examples} Examples}

\subsection{Simulation studies}
We carry out simulations to compare the use of the  method described in this paper  
with the method COBS proposed by  \shortciteN{He1999}. COBS  estimates both constrained and unconstrained quantile curves using B-spline smoothing and is available as 
an R package. Here we use the unconstrained case as a fully automated procedure, where both the smoothing parameter   and the 
selection of knots is carried out  according to either the AIC or the BIC criterion.

We consider simulated datasets that correspond to the three examples described bellow, these examples are adapted from some well known examples in the curve fitting literature, see e.g. \shortciteN{smith+kohn96}, \shortciteN{denison+MS98} and \shortciteN{dimatteo+gk01}.
\begin{description}
\item[\it Example 1:] 
Here the curve takes the form
$$
f(x)=\phi(x,0.15,0.05^2)/4+\phi(x,0.6,0.2^2)/4,\quad x\in[0,1],
$$
where $\phi(x,\mu,\sigma^2)$ denotes the value at $x$ of the normal density with mean $\mu$ and variance $\sigma^2$. $n=200$ data 
points $x$ are sampled from the Uniform distribution $U(0,1)$. The noise $\epsilon$ 
is added to the data, they corresponds to a Gamma distribution 
${\cal G}a(1,4)$ with shape parameter 1 and rate parameter 4 that is translated by  -$0.175$ (so that the median of this noise distribution is 
approximatively 0).

\item[\it Example 2:] Here
the curve takes the form
$$
f(x)=\sin(2x) + 2\exp(-16x^2),  \quad x \in [-2,2].
$$
and is evaluated at $n=201$ regularly spaced grid points. This function is first rescaled so that the support is on the unit interval. The noise $\epsilon$ added to the data is simulated in the same way as in the first example.

\item[\it Example 3:] In this example the curve is given by, 
$$
f(x)=\sin(x) + 2\exp(-30x^2),  \quad x \in [-2,2],
$$
and the data points $x$ correspond to $n=201$ regularly spaced grid points. As in the previous example the function is rescaled on the unit interval for $x$ and the same distribution for noise $\epsilon$ is used for the data.
\end{description}

\noindent To compare the different methods we use  the mean squared error (MSE) as a measure of goodness of fit, given by
$$\mbox{MSE}=\frac{1}{n}\sum_{i=1}^n \{\hat{f}(x_i)-f(x_i)\}^2$$
where $f$ is the true median regression function and $\hat{f}$ is the estimated function. Since the COBS algorithm computes the median curve with quadratic
 (or linear) splines we consider hereafter the case $P=2$.
 
For the  prior specifications of each example,  we set  $\lambda = 3$ and $L = 10$ for the truncated Poisson prior. Results are largely insensitive to values of 
$\lambda$ around this range and the maximum number of knots allowed $L$ is chosen to be large enough to not affect the simulation results here. For these 
examples we consider the situation where there is no prior information on the knot locations and chose the intervals $I_k$ to correspond to the ranges given by 
every $n_x$ sorted $x$ values. We found that $n_x=5$ was sufficient to provide a good fit in each of the three examples. We use a B-spline basis to formulate 
the $X_{z,\gamma}$ matrix, as in \shortciteN{dimatteo+gk01}, to  avoid numerical instability (see e.g. \shortciteNP{ruppertwc03}). 

\begin{figure}[htb]\begin{center}
\begin{tabular}{c}
\subfigure[Example 1]
{
\includegraphics[height=9cm, angle=-90]{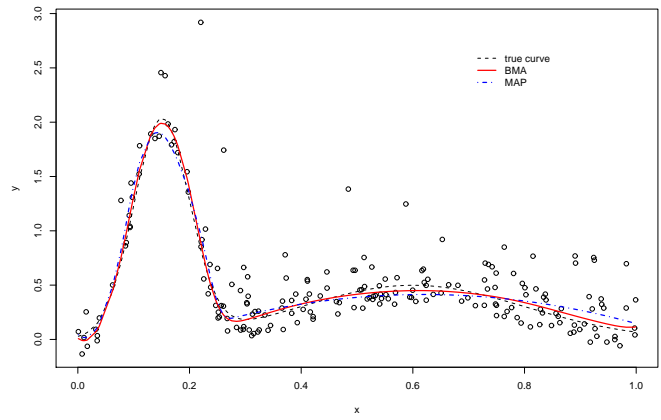}
}
\\
\subfigure[Example 2]
{
\includegraphics[height=9cm, angle=-90]{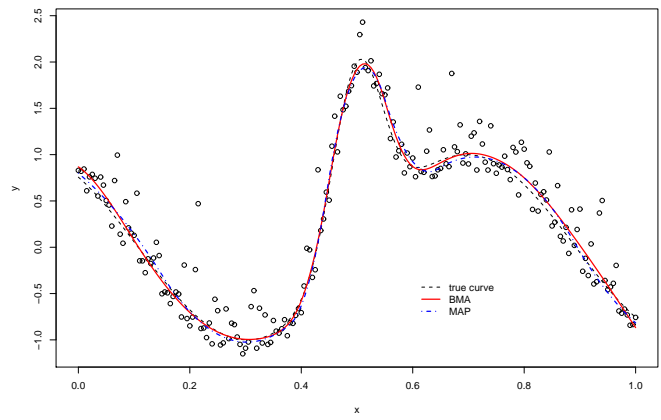}
}
\\
\subfigure[Example 3]
{
\includegraphics[height=9cm, angle=-90]{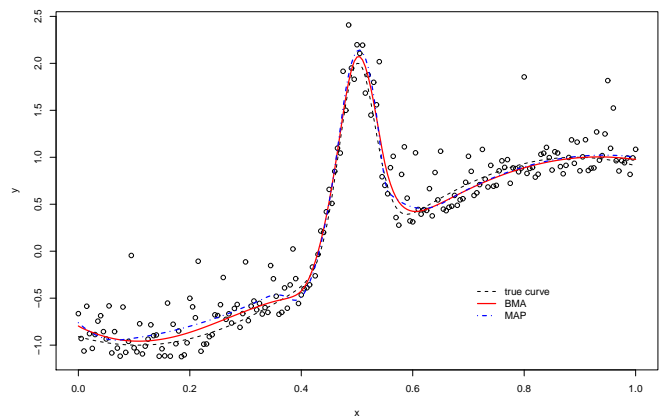}
}
\end{tabular}
\caption{{\small Estimated curves for the three simulated examples.}}
\label{fig:examplesCurves}
\end{center}
\end{figure}

The  computation of all three examples started with an arbitrary set of initial values generated from the prior distributions. 
We first ran the algorithm 500 iterations for adaptive tuning then fixing the scaling parameters of the Random-Walk Metropolis Hastings algorithm at the final value of the tuning run,  we then ran a burn-in of 500 iterations, followed by 1,500 recorded iterations. 
Each iteration involves an update of 20 $z$ update steps for each $\gamma$ update step. To assess convergence, we monitored the trace plots of each model parameters as well as posterior values. We also ran much longer chains of 10,000 iterations and found the results to be similar in terms of MSE calculations. 
See Figure \ref{fig:examplesCurves}  for the fitted functions of the three examples using our method with the BMA estimate and with the MAP estimate. 

\begin{table}
$$
\begin{array}{|c|c|c|c|c|}
\hline
    & \mbox{ BMA } & \mbox{ MAP } & \mbox{ COBS } & \mbox{ COBS } \\
    &                          &                           &     \mbox{AIC}              &    \mbox{BIC}                   \\
\hline 
\mbox{Example 1} & 0.0032 &  0.0055 &  0.0052 & 0.0075 \\
   & (0.0017) & (0.0021) &  (0.0033)  & (0.0069)\\
\hline
\mbox{Example 2} &  0.0040 &  0.0067 & 0.0060 & 0.0065 \\
   & (0.0025) & (0.0038) & ( 0.0029 ) & (0.0026)\\
\hline
\mbox{Example 3} & 0.0036  &  0.0056 &  0.0084  & 0.0139 \\
   & (0.0018) & (0.0025) &    ( 0.0028) & (0.0044)\\
\hline   
\end{array}
$$
\caption{{\small Mean MSEs with estimated standard errors in brackets based on 50 samples obtained using the Bayesian model averaging (BMA), the  
maximum a posteriori (MAP) and the COBS algorithm (with the AIC criterion and with the BIC criterion). }}
\label{tableEx}
\end{table}

\begin{figure}[htb]
\begin{center}
\begin{tabular}{c}
\subfigure[Example 1]
{
\includegraphics[height=9cm, angle=-90]{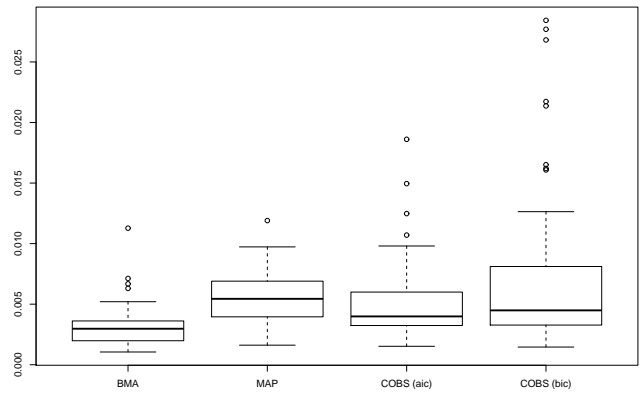}
}
\\
\subfigure[Example 2]
{
\includegraphics[height=9cm, angle=-90]{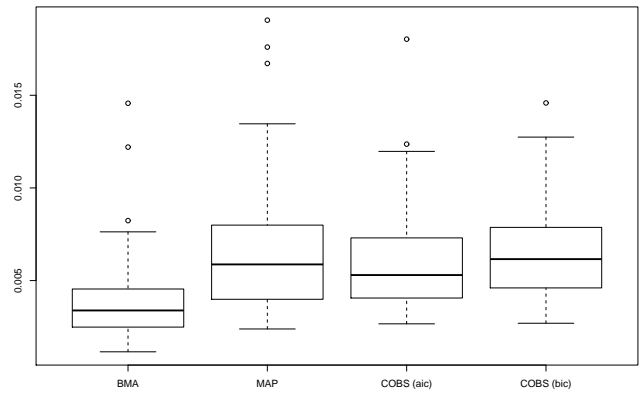}
}
\\
\subfigure[Example 3]
{
\includegraphics[height=9cm, angle=-90]{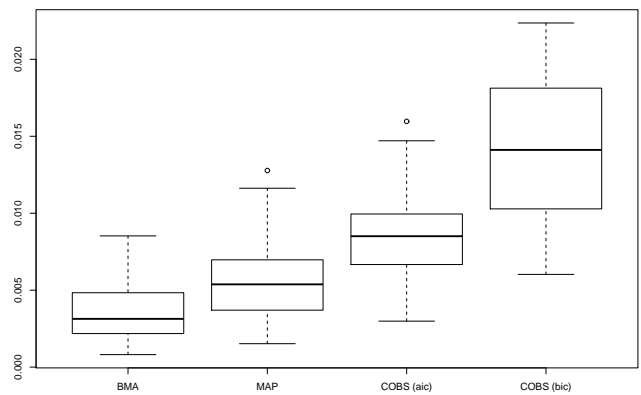}
}
\end{tabular}
\caption{{\small Boxplots for the MSEs corresponding to the three simulated examples.}}
\label{fig:examplesBoxplots}
\end{center}
\end{figure}

For each of the three examples the BMA, MAP and COBS (with the AIC or the BIC criterion) estimates are calculated over 50 randomly generated datasets. The mean and standard deviation of the  MSEs are presented in Table \ref{tableEx}, the corresponding boxplots are given in Figure \ref{fig:examplesBoxplots}. 
On the whole the method presented in this paper performs well compared to  COBS, especially on the datasets corresponding to Example 3. 
On the three types of datasets that are  considered here, the BMA estimates seem to be more accurate than the MAP estimates.

\subsection{Motorcycle data set}
We consider a reference dataset, the motorcycle data,  studied in the context of quantile regression for example in \shortciteN{Koenker2005} or in 
\shortciteN{ChenYu2009}. These data are analyzed in \shortciteN{Silverman1985} and  contain experimental measurements of the acceleration of the head of 
  a test dummy  (expressed in $g$, acceleration due to gravity) as a function of time in the first moments after an impact (the time is expressed in $ms$). The dataset is challenging for quantile regression as the the values and the variability of the response  vary dramatically with the independent variable. 

We fit to these data the quantile regression curves corresponding to $p=0.25,0.5$ and $0.75$. The prior settings are esentially the same as the ones already described in the simulation studies, except here  we set  $\lambda=5$ and $L=15$ for the truncated Poisson prior. For the MCMC computation of the curves we started with an arbitrary set of initial values generated from the prior distributions. 
Again we used the first  500 iterations for adaptive tuning then we ran a burn-in of 500 iterations  followed by 3,500 recorded iterations, where each iteration involves an update of 20 $z$ update steps for each $\gamma$ update step. 

\begin{figure}[h]
\begin{center}
\includegraphics[height=9cm, angle=-90]{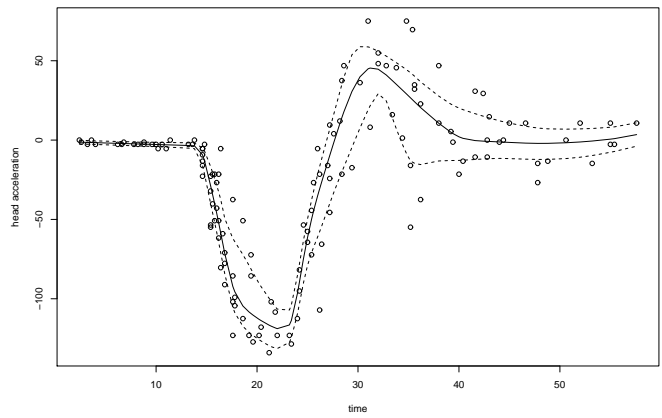}
\caption{{\small Motor cycle data set  ; estimated quartile regression curves by BA approach for $P=1$ (p=0.5: solid line ; p=0.25 and p=0.75: dotted lines)}}
\label{figmotorcycle}
\end{center}
\end{figure}

We give in Figure \ref{figmotorcycle} 
the quantile curves corresponding to linear splines $P=1$.
The results appear quite satisfactory as the quantile curves are not crossing each other, even in the region beyond 50 millisecond where the data are sparse. 
The changes in the variability of the  acceleration over time has been captured well by the  fitted conditional quantile curves,  as they are very close to each other for the first few milliseconds  then diverge after the crash.


\section{\label{addMod} Quantile regression  for additive models}
\subsection{Introduction}
When several potential predictors for the response variable are of interest, a standard procedure to avoid the so-called ``curse of dimensionality" is to use an additive model (\citeNP{hastie+t90}) where the response is modeled as a sum of functions of the predictors. In the context of quantile regression, if $Y$  denotes the real-valued response variable and if now ${\bf X}=(X^1,...,X^d)$ denotes a vector of $d$ predictors, the $p$-th quantile of the conditional distribution of $Y$ given $\{{\bf X}={\bf x}\}$ is modeled as
\begin{eqnarray}
f_p({\bf x}) & = & \sum_{j=1}^d f_p^j(x^j).
\label{addModQuant}
\end{eqnarray}
See \shortciteN{YuLu2004} for an inference on the additive quantile regression model by a kernel-weighted local linear fitting and see 
 \shortciteN{YueRue2011}  for a Bayesian inference with either a MCMC algorithm or using INLA. 
 
If we use  spline functions to model the different curves $f_p^1(x^1), ..., f_p^d(x^d)$ it is still possible to use  the linear model (\ref{eqn:varsele})  
with the difference that  the design matrix $X_{\gamma}$ is now made up of the columns of the individual design matrices corresponding to (\ref{designMatrix}), 
with a single intercept term for identifiability. 
Thus inference on the additive quantile regression model can be performed via the same methodology and 
algorithm described in the previous sections. We consider below the study of a real dataset that involves additive quantile regression.

\subsection{Analysis of the Boston housing dataset}
We revisit  the  so-called Boston house price data  available in the R package MASS. 
This dataset has been  originally studied in  \shortciteN{HarrisonRubinfeld78}. 
The full dataset  consists of the median value of owner-occupied homes in 506 census tracts in the Boston Standard Metropolitan Statistical Area in 1970 
 along with 13  various sociodemographic variables.
This dataset has been analyzed in many statistical papers including \shortciteN{OpsomerRuppert1998}, who used an additive model for mean regression, 
and \shortciteN{YuLu2004}, who proposed an additive quantile regression model by a kernel-weighted local linear fitting. As in these two references we consider 
 the median values of the owner-occupied homes (in \$1000s) as the dependent variable and four covariates  given by \\
 
 RM = average number of rooms per house in the area, 
 
 TAX = full property tax rate (\$/\$10,000),
 
 PTRATIO = pupil/teacher  ratio by town school distric,
 
 LSTAT = the percentage of the population having lower economic status in the area.\\
 
\begin{figure}[h]
\begin{center}
\includegraphics[height=15cm, angle=-90]{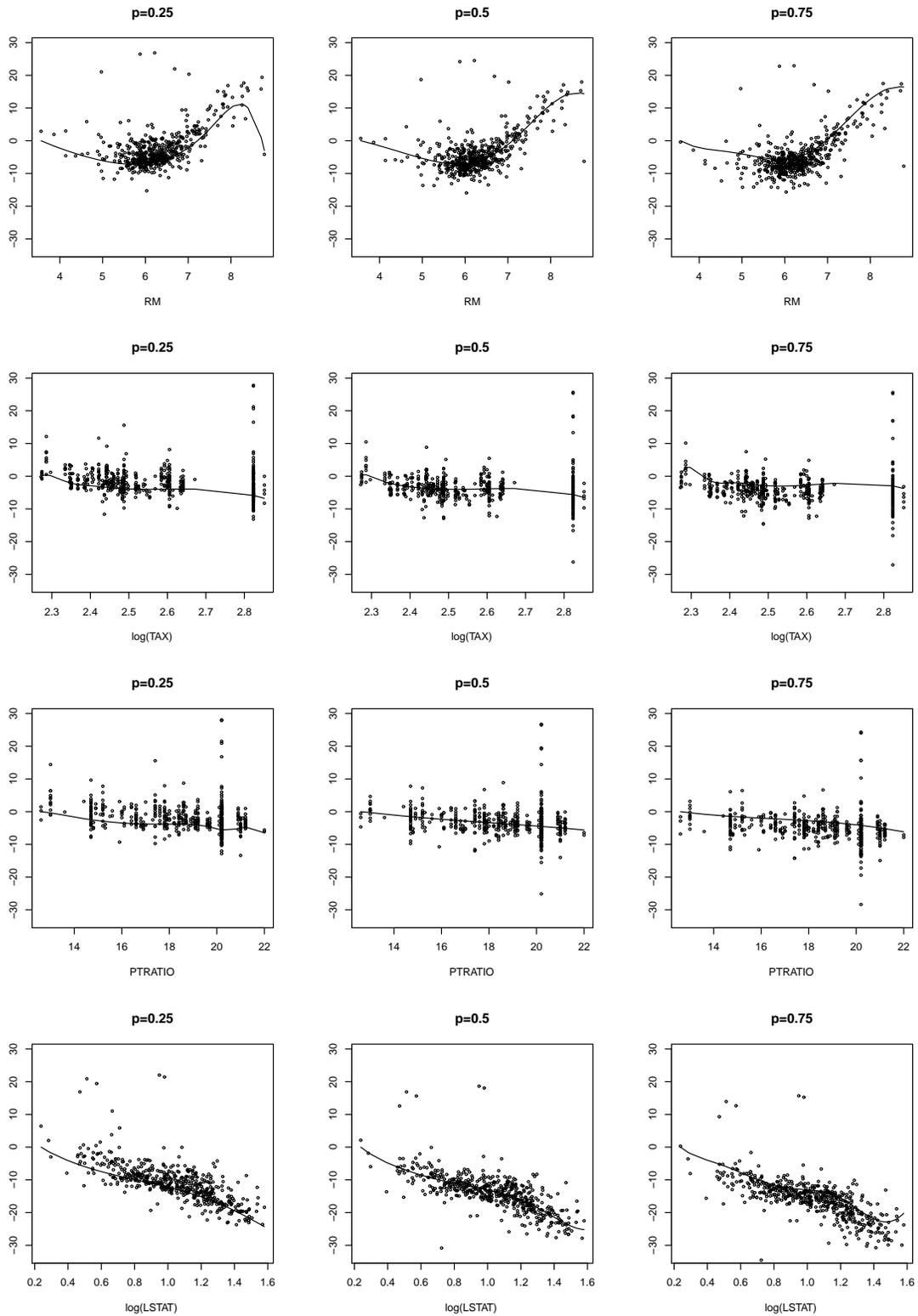}
\caption{{\small Boston housing dataset ;  fitted quantile curves $p=0.25,0.5,0.75$, $P=3$, for the four variables that have been considered. One each figure the datapoints represented correspond to the original data minus the effect of all the other variables (and the constant term).}}
\label{figBostonHousing}
\end{center}
\end{figure}

 \noindent As noticed in \shortciteN{YuLu2004} these data are  suitable for a quantile regression analysis since the response is a median price in a given area and the variables RM and LSTAT are highly skewed. More precisely we consider the additive model where the $p$-th quantile of the conditional distribution of the response is given by
\begin{eqnarray*}
f_p({\bf x}) & = & \alpha_0+f_p^1(\mbox{RM})+f_p^2(\log(\mbox{TAX}))+f_p^3(\mbox{PTRATIO})+f_p^4(\log(\mbox{LSTAT})).
\end{eqnarray*}
We fit to these data the $p$-th quantile regression curves corresponding to cubic splines ($P=3$) at the quantile levels  $p=0.25,0.5$ and $0.75$. 
For the prior settings  we took $\lambda=5$ and $L=8$ for the truncated Poisson prior. 
For each predictor we set the intervals $I_k$ to be 10 equally sized partition sets  over the range of the variable.  Excluding the possibility of knots in the first and 
the last intervals,  we get  $K_{max}=8$ for each variable.  
For the MCMC computation of the curves we started with a random  set of initial values generated from the prior distributions.  
We first ran the algorithm 500 iterations for adaptive tuning then we ran a burn-in of 500 iterations, followed by 4,000 recorded iterations, where each iteration involves an update of 20 $z$ update steps for each $\gamma$ update step.  We present in Figure \ref{figBostonHousing}  the different estimated curves. We plotted on the same graphs the datapoints  corresponding to the original data minus the effect of all the other variables and the constant term. The fact that the values of $\log($TAX$)$  are not well dispersed over their range and the  presence of a few outliers in the dataset did not seem  to be a problem for our method.

Our results appear consistent with the results provided in  the quoted previous analyses. Briefly, the variables RM and LSTAT appear as the most important covariates. If  the contribution of LSTAT look similar for the three quantiles levels, the contribution of RM  looks slightly more important for the upper quantile level $p=0.75$. The variable TAX has a contribution relatively more important for the lower quantile level $p=0.25$. Finally the Figure \ref{figBostonHousing} suggests a linear contribution of the variable PTRATIO, especially for $p=0.5$ and for $p=0.75$.

\section{\label{crossing} Noncrossing quantile regression curves}
\begin{figure}[htb]
\begin{center}
\begin{tabular}{c}
\subfigure[Initially estimated curves]
{
\includegraphics[height=9cm, angle=-90]{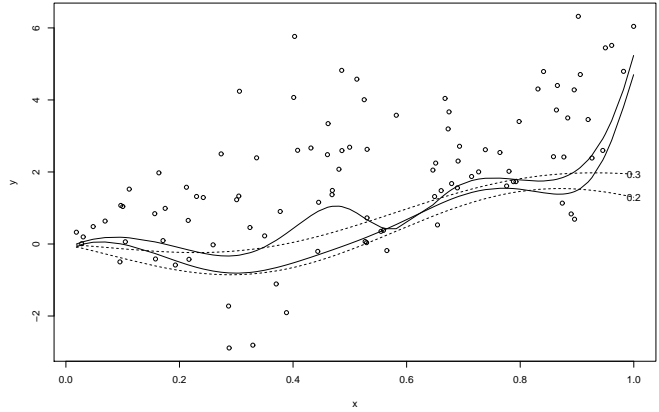}
}
\\
\subfigure[Corrected curves]
{
\includegraphics[height=9cm, angle=-90]{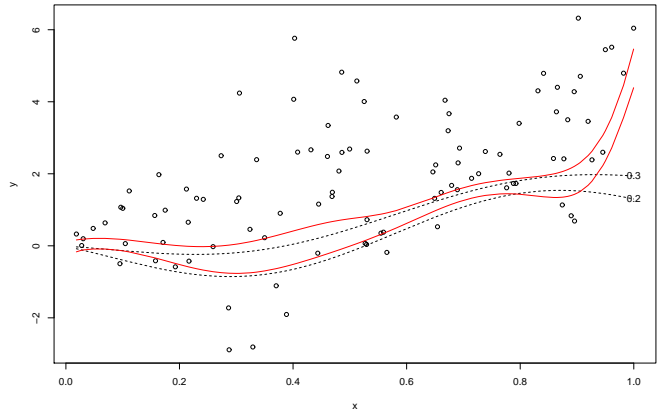}
}
\end{tabular}
\caption{{\small Curve crossing example. The dotted lines represent the true quantile curves for $p=0.2$ and $p=0.3$. The solid lines represent (a) 
the quantiles curves that have been estimated separately (b) the corrected estimated quantile curves with respect to the new substitution likelihood.}}
\label{fig:examplesCross}
\end{center}
\end{figure}

One known problem when using quantile regression for multiple percentiles is that the  quantile curves that are estimated separately can cross, which is impossible. 
See for example the Figure \ref{fig:examplesCross} (a) where, partly due to the relatively small size of the dataset and the complex conditional distribution of the response variable,  the two estimated quantile curves for $p=0.2$ and $p=0.3$ are crossing around the value $x=0.6$. 

The treatment of noncrossing quantile regression curves is  difficult and several attempts to circumvent this problem have been proposed in different settings, see {\it e.g.}  the references in  \shortciteN{Yu2003} and  in \shortciteN{Koenker2005} or, for a more recent development in this area, see {\it e.g.}  \shortciteN{ReichFuentesDunson2011}. 
In particular,  \shortciteN{Bondell2010}  proposed a solution to this problem by considering a generalization of the criterion (\ref{checkfunction}) to the case of simultaneous inference on several quantile curves. For clarity we suppose hereafter that we are interested in the fitting of two quantile curves corresponding to quantile levels $p_1$ and $p_2$, with $p_1<p_2$.  \shortciteN{Bondell2010} gave a solution to the minimization under the constraint $f_{p_1}(.)<f_{p_2}(.)$ of the expression 
\begin{eqnarray}
\sum_{j=1}^2   \left\{    \sum_{i=1}^n \rho_{p_j}(y_i-f_{p_j}(x_i))  \right\}
\label{checkfunctionBondell}
\end{eqnarray}
plus a penalty term corresponding to smoothing. 
An alternative approach  described in \shortciteN{DunsonTaylor2005} uses a so-called ``substitution likelihood'' that does not correspond to the distribution of the data given the unknown curves but yields a valid uncertainty. The substitution likelihood that they considered corresponds to the multinomial weights
\begin{eqnarray}
s(f_{p_1},f_{p_2}|Y) & = & \frac{n!}{u_1 ! u_2 ! u_3 !} p_1^{u_1} (p_2-p_1)^{u_2} (1-p_2)^{u_3} I_{\{   f_{p_1} <f_{p_2}   \} }
\label{substitutionDunson}
\end{eqnarray}
where $u_1$ represents the number of datapoints below the curve $f_1$, where $u_2$ represents the number of datapoints between the two curves and where  $u_3$ is the number of datapoints above the curve $f_2$. They gave conditions on the prior for the ``pseudo-posterior'' $\pi( f_{p_1},f_{p_2} | Y   ) \propto   s(f_{p_1},f_{p_2} |Y) \pi(  f_{p_1},f_{p_2}  )$   to be proper and proposed a MCMC algorithm for (pseudo-)posterior computation in the case of linear quantiles. \\

Here we propose a new method to postprocess the MCMC samples obtained from separate quantile  regression curve fitting.  We denote by $\theta_p=(\beta_{z,\gamma},\sigma,z,\gamma,W,c)$ the full set of unknown parameters for the $p$-th quantile regression curve model (\ref{condYquantilebis}).   Let $\theta_{p_1}$, $\theta_{p_2}$,  be the parameters corresponding to the quantile regression curve for the quantile levels $p_1$ and  
  $p_2$ respectively, $p_1<p_2$. We consider a new substitution likelihood  of the form 
\begin{eqnarray}
s(\theta_{p_1},\theta_{p_2}|Y) & = &  L  (\theta_{p_1}|Y) L (\theta_{p_2}|Y)  I_{\{   f_{p_1}(x|\theta_{p_1} )< f_{p_2}(x|\theta_{p_2})  \} }
\label{newsubstitution}
\end{eqnarray}
where $L  (\theta_{p_1}|Y) $ and $L  (\theta_{p_2}|Y) $ denotes the two likelihood functions  for quantile levels  $p_1$ and $p_2$ given by the conditional distribution (\ref{condYquantilebis}). The indicator function takes the value one if $f_{p_1}(x|\theta_{p_1} ) < f_{p_2}(x|\theta_{p_2})$ for all $x$ and zero otherwise, here the function $f_{p}(x|\theta_{p} )$ is evaluated according to Equation (\ref{eqn:spline}) with parameters $\theta_p$. It is not hard to see that the maximizer of this substitution likelihood  is the maximizer of  (\ref{checkfunctionBondell}).  Moreover, if we take independent priors $\pi(\theta_{p_1})$ and $\pi(\theta_{p_2})$ on the two sets of parameters,  then the corresponding quasi-posterior is simply 
\begin{eqnarray}
\pi(\theta_{p_1},\theta_{p_2}|Y)  & \propto &  s(\theta_{p_1},\theta_{p_2}|Y) \pi(\theta_{p_1}) \pi(\theta_{p_2}) , \nonumber \\
& \propto &  \pi (\theta_{p_1}|Y) \pi (\theta_{p_2}|Y)   I_{\{   f_{p_1}(x|\theta_{p_1} )< f_{p_2}(x|\theta_{p_2} )  \} }.
\label{newpost}
\end{eqnarray}
Given samples from the distribution  $\pi (\theta_{p_1}|Y) \otimes \pi (\theta_{p_2}|Y)  $ an importance sampling argument can be used to reweight the samples according to this quasi-posterior.

In practice, MCMC samples obtained from separate posterior explorations of $\pi (\theta_{p_1} |Y)$ and of $\pi (\theta_{p_2} |Y)  $ can be combined to form the  new estimate of the the curve $f_{p_1}(x)$ by
\begin{eqnarray}
\mathbb{E}_{\pi(\theta_{p_1} ,\theta_{p_2} |Y)}[f_{p_1}(x|\theta_{p_1} )]  & \approx &  \frac{ \sum_t  f_{p_1}(x|\theta_{p_1}^t)   I_{\{   f_{p_1}(x|\theta_{p_1}^t )< f_{p_2}(x|\theta_{p_2}^t)  \} } }   {   \sum_t   I_{\{   f_{p_1}(x|\theta_{p_1}^t )< f_{p_2}(x|\theta_{p_2}^t)  \} }  }.
\label{approxnewpost}
\end{eqnarray}
When the constraint above excludes too many samples this estimator will be unreliable, in this case more MCMC samples will be required. A computationally cheap way to obtain more samples is to consider all combinations of the two MCMC samples.

Figure \ref{fig:examplesCross} (b) shows the corrected curves from the estimator in (\ref{approxnewpost}), using  all the possible combinations of the two MCMC samples, each of size 2,000. To evaluate the curves  we use here the plug-in estimator (\ref{postBetaQuant}) for $\beta_{z,\gamma}$. Finally the constraint on the curves is checked at every observed values of $x$.

The strength of the above approach is that it is very easy to apply, and can be used on any  posterior samples from separate quantile curves. An obvious draw back is that in some cases, when for example $p_1$ and $p_2$ are very close, the number of samples satisfying the constraint can be extremely low.

\section{\label{discussion} Conclusion}
In this article, we have provided a procedure for Bayesian inference on quantile curve fitting. We focused on the use of regression splines with unknown number of knots and location to obtain smooth curves.  We have seen that, within an auxiliary variable framework, a scale mixture of normals representation for the asymmetric Laplace distribution together  with appropriate prior specifications makes it possible to integrate out the  regression and the variance parameters analytically. This facilitates a simple Metropolis- Hastings within Gibbs sampler for simulation from the posterior distribution of interest. 
The proposed algorithm is fully automated with the inclusion of an automatic tuning step, which optimally tunes the Random-Walk Metropolis-Hastings scaling parameters.  We have shown that our method performs well on several types of datasets. We have also shown that the proposed framework can be trivially extended to inference on additive models. Finally we have proposed and discussed a simple and general procedure that postprocesses  MCMC samples to obtain  noncrossing 
quantile regression curves.

\bibliographystyle{chicago}
\bibliography{QuantReg}

\section*{Appendix A}

\subsection*{The marginal posterior}
The full joint posterior distribution of the parameters  is 
\begin{eqnarray*}
\pi(   \beta_{z,\gamma},z, \sigma, \gamma,W,c|Y ) & \propto & f(Y| X_{z, \gamma},\beta_{z,\gamma},z, \sigma, \gamma ,W) \pi(W|\sigma) \pi( \beta_{z,\gamma},z, \sigma, \gamma|W ) \pi(c) ,\\
 & \propto &  f(Y| X_{z, \gamma},\beta_{z,\gamma},z, \sigma, \gamma ,W)  \pi(W|\sigma) \pi_{\beta_{z,\gamma}}(\beta_{z,\gamma}|z, \sigma, \gamma,W) \\
 & & \qquad \times \pi_{\sigma}(\sigma)\pi_z(z)\pi_{\gamma}(\gamma) \pi(c).
\end{eqnarray*}
With $f(Y| X_{z, \gamma},\beta_{z,\gamma},z, \sigma, \gamma ,W)  $ and   $\pi_{\beta_{z,\gamma}}(\beta_{z,\gamma}|z, \sigma, \gamma,W)$   given 
by the two Gaussian distributions (\ref{condYquantilebis}) and (\ref{unitinfoQuant}) the parameter $ \beta_{z,\gamma} $ is easily integrated out using 
classical results of Bayesian linear models. We get 
\begin{eqnarray*}
\pi(   z, \sigma, \gamma,W,c|Y ) & \propto &    \left(\frac{1}{c+1} \right)^{\frac{|z|+P+1}{2}} \frac{ \pi_z(z) \pi_{\gamma}(\gamma)}{  \sqrt{ \pi_{i=1}^n w_i }  }  \left( \frac{1}{\sigma} \right)^{1+\frac{3n}{2}} e^{-\frac{1}{\sigma}\left\{ \frac{p(1-p)}{4}S_{z,\gamma,W}(Y) +\sum_{i=1}^n w_i \right\}} \pi(c)
\end{eqnarray*}
where 
\begin{eqnarray*}
S_{z,\gamma,W}(Y) & = & Y_{(W)}'W^{-1}Y_{(W)}-\frac{c}{c+1}Y_{(W)}'W^{-1}X_{z,\gamma}(X'_{z,\gamma}W^{-1} X_{z,\gamma})^{-1}X'_{z,\gamma}W^{-1}Y_{(W)}
\end{eqnarray*}
with
\begin{eqnarray*}
Y_{(W)} & = & Y-\frac{(1-2p)}{p(1-p)}W {\bf 1}_n.
\end{eqnarray*}
Then the parameter $\sigma$ can be also integrated out and we get
\begin{eqnarray*}
 \pi(z, \gamma,W,c |Y) & \propto &     \left(\frac{1}{c+1} \right)^{\frac{|z|+P+1}{2}} \frac{ \pi_z(z) \pi_{\gamma}(\gamma)}{  \sqrt{ \pi_{i=1}^n w_i }  }    \left(  \frac{1}{ \frac{p(1-p)}{4}S_{z,\gamma,W,c}(Y)  +\sum_{i=1}^n w_i } \right)^{\frac{3n}{2}}  \pi(c).
\end{eqnarray*}

\section*{Appendix B}
\subsection*{Automatic tuning algorithm}
Here we provide the algorithm to optimally search for the tuning parameters $\sigma^2_{i}, i=1,\ldots, n$, and $\sigma^2_*$. The algorithm runs within the main MCMC algorithm given in Section \ref{sec:alg}. Tuning will only apply to the Update $W$ and Update $c$ steps. For the update of each of the parameters $w_i, i=1,\ldots, n$ and $c$ do:\\

\noindent{\bf Initialisation:}
For the iteration $t=1$ of the algorithm  initialise the  scaling parameter $\sigma^* = \sigma^1=1$, where $\sigma^1$ corresponds to $\sigma_{i}$ and $\sigma_*$ when updating the parameters $w_i$, $i=1,\ldots,n$ and $c$ respectively.
Set $p^*=0.44$ and initialise $j=0$ . The value of $p^*$ corresponds to the optimal acceptance probability for a univariate Random-Walk Metropolis-Hastings algorithm (\shortciteNP{roberts+r01}).\\

\noindent{\bf Tuning:}  Set $j=j+1$; update the parameters according to either Update $W$ or update $c$, and obtain the corresponding acceptance probability $\alpha$ as in Section \ref{sec:alg}. \\

\underline{{\bf {\em update scaling:}}} if $j<20$ set  $\sigma^{t+1}=\sigma^{t}$, else set
$$
\sigma^{t+1}=\left\{
\begin{array}{ll}
\sigma^t+\kappa (1-p^*)/j &\mbox{if } U < \alpha\\
\sigma^t-\kappa p^*/j &\mbox{if } U >  \alpha
\end{array}\right.
$$
where $\kappa=\sigma^t/\{p^*(1-p^*)\}$ and $U\sim U(0,1)$.\\

\underline{{\bf {\em restart the algorithm:}} }If $t<100$, and either $\sigma^{t+1} > 3\sigma^*$ or $\sigma^{t+1} < \sigma^*/3$, restart the algorithm, setting $\sigma^*=\sigma^{t+1}$ and $j=0$. Note we do not restart the algorithm again if the total number of restarts exceeds 5.
\\

\underline{{\bf {\em Increment loop: }}} Set $t=t+1$. Go back to the beginning unless $t$ exceeds some prespecified number of iterations $nTune$.
\\

\noindent It is easy to monitor the changes in $\sigma^t$ in order to determine the number of tuning iterations $nTune$ to achieve stability. In practice, we run the first $nTune$ iterations of the algorithm in Section \ref{sec:alg} with automatic tuning, and then start the main part of MCMC as usual with the scaling parameters fixed at the value $\sigma^{nTune}$.


\end{document}